\documentclass[a4paper]{article}

\usepackage[english]{babel}
\usepackage[utf8]{inputenc}
\usepackage{amsmath}
\usepackage{graphicx}
\usepackage[colorinlistoftodos]{todonotes}
\usepackage{authblk}
\usepackage{color}
\usepackage{ulem}

\title{\textcolor{blue}{A functional approach to the next to eikonal approximation of high energy gravitational scattering} }

\author[1]{ A.R. Fazio}
\affil[1]{ \small \textsl{ Departamento de F\'{i}sica, Universidad Nacional de Colombia, Ciudad Universitaria, Bogot\'{a} D.C., Colombia}}
\author[2]{ E.A. Reyes R.}
\affil[2]{ \small \textsl{ Universidad de Pamplona (UDP), Pamplona - Norte de Santander, Colombia }}
\date{}

\begin{document}
\maketitle

\begin{abstract}
The Fradkin-Schwinger functional methods to represent a Green function in an external gravitational field are used to study the eikonal and the next-to-eikonal limit, including the nonlinear gravitational interactions, of the scattering amplitudes of an ultra-relativistic scalar particle on a static super-massive scalar target in the nearly forward limit. The functional approach confirms the exponentiation of the leading eikonal which also applies to the first non-leading power in the energy of the light particle, moreover includes the interaction at impact parameter much larger than the Schwarzschild radius associated with the center of mass energy in the ultra-relativistic limit.
\end{abstract}

\section{Introduction}

The high-energy behavior of scattering processes in perturbative quantum gravity of a very light scalar by a heavy scalar \cite{Hooft},\cite{Hooft1},\cite{Verlinde} is considered in the eikonal and next-to-eikonal approximation \cite{wallace} for the large impact parameter and the consequent small scattering angle limit. These approximations schemes notably go beyond the finite order perturbation theory \cite{Glauber},\cite{Fried}, allowing to sum the logarithmic divergent amplitudes in soft limit which have ladder like structure \cite{wein},\cite{akhouri},\cite{beneke}. The general idea of the eikonal limit is that at very high-energies the de Broglie wave-length of the incident particle of a scattering process is small compared with the spatial variations in the target, so that the ``geometrical optics"-limit might be reasonable as a first approximation in which it is possible to recover the classical result of the exponentiation of the leading phase \cite{Levy}. That phase depends upon the mass dimension of the coupling constant which is for gravitational interactions inverse of squared energy \cite{dietz} and it is a fundamental infrared dominated physical quantity \cite{Giddings}. By using and developing scattering amplitudes techniques \cite{Elvang} more kinematic regimes and observables have been explored and also in generalized theories with gravity compared with classical gravitational scattering of massive objects untill two loop order \cite{Bern}. However, the approach of scattering amplitudes in perturbative quantum gravity to be compared with classical dynamics for the post Minkowskian expansion needs an expansion in the small scattering angle \cite{damour} in which the amplitude can be resummed by exponentiating a suitably defined eikonal phase. As we will see in our analysis, there are subtleties about which perturbative diagrams do exponentiate and possibly, by the resummation approach used in this paper, the extension to two loops could be afforded in order to compare with the results from the scattering amplitudes. The extraction of classical results has been also studied in \cite{kos1}, \cite{kos2} and the eikonalization has been applied in that context \cite{monteiro}.
In the reference \cite{Sterman} a detailed analysis of the expansion around the eikonal limit directly in perturbation theory for the above mentioned scattering is provided and the diagrammatic self-consistency of this expansion is verified with the derivation of the next-to-leading corrections in arbitrary dimensions. The first derivation of the eikonal gravitational phase for the process in question can be found in \cite{kabat}. The next-to-leading correction has been also studied intensively \cite{DIV},\cite{CD},\cite{Luna},\cite{Collado},\cite{Naculich}. In our article we provide by the functional methods of quantum field theory the summation of diagrams corresponding to the treatment of the leading and the next-to-leading corrections of the high-energy scattering amplitude of the considered process. In the framework of the functional integral, the Green function representation in an external field proposed by Fradkin in \cite{Fradkin1}, \cite{Fradkin2} allows to perform calculations in a compact form with easy combinatorial counting based on the Wick’s contractions in the summations of ladder diagrams. Two nucleons scattering in quantum gravity was also analyzed by the so called Fradkin’s modified perturbation theory \cite{xuan}, however, our approach is based on the Feynman path integral and takes into account in the sum of the considered diagrams the full Feynman denominators like $(p+\sum k_i)^2-m^2 +i\epsilon$, by expanding in the correlation terms $k_i\cdot k_j$ for the internal momenta. The quantum gravity regime is the appropriate to discuss the ultrahigh energy gravitational scattering in the eikonal approximation because it is characterized by the large impact parameter where the massless fields dominate. They cannot be strings modes because in \cite{gross} has been proved that when the momentum transfers reach the string scale strings effects appear to be subdominant to higher loop gravitational processes approximated via eikonal. The high-energy scattering of massless closed strings from a stack of a given number of D-branes in Minkowski space is characterized also by the dominance of gravity when the impact parameter is large with respect to the  Schwarzschild radius associated with the center of mass energy~\cite{DIV}. The eikonalization procedure of \cite{DIV} is based on the use of Schwinger proper time, which in that context is expressed in terms of the strings and D-brane parameters in perturbation theory. In our approach the use of Schwinger proper time will allow an efficient eikonalization procedure with a suitable extension at next-to-eikonal approximation. 

The paper is organized as follows. In section \ref{UNO} we fix the kinematics of the high energy gravitational scattering process of the spinless projectile in the static limit of the spinless target. In section \ref{DUE} the path integral description of the amplitude by semiclassical gravity is provided. In section \ref{TRE} the Fradkin’s functional representation of the two-point connected Green function in a linearized background is derived in details. That representation allows to take into account all emitted gravitons along the propagating line of the radiating particle with the truncation to linear interaction in the graviton field. In section \ref{QUATTRO} the eikonal limit of the two point Green function in the given external gravitational field is computed. The eikonal limit of the scattering amplitude of our process is computed by showing crucial cancellations at next-to-power in the transfer momentum and also by showing the large impact parameters dominance for the process in the eikonal regime. In section \ref{CINQUE} the deviation from the straight line is considered in our functional approach obtaining zero at the next to leading power in four dimensions. In section \ref{SEI} the non-linear gravitational interactions are included for the contribution at the next to leading power. Section \ref{SETTE} is for conclusions and research perspectives to apply the functional methods to problems in the next-to-eikonal approach proposed in the current literature. 

\section{Kinematics\label{UNO}}

We will investigate the small angle gravitational scattering of an ultraviolet light (massless) scalar particle of energy $E_\phi$ off a very heavy particle of mass $M_\sigma$, also chosen to be scalar. The specific kinematics is
\begin{equation}
p+q\rightarrow p’+q’\,\,\,\,\,\,\, p^2=p’^2=0\,\,\,\,\,\,\, q^2=q’^2=M_\sigma^2
\label{process}
\end{equation}
and we indicate by $\Delta\equiv\sqrt{-(p-p’)^2}<<E_\phi=p^0<< M_\sigma$. The momenta of the incoming and outgoing scalar particles, $p$ and $p’$, are much larger than the transferred momentum $\Delta^\mu=p^{\prime\mu}-p^{\mu}$. We will take $q$ and $q’$ to be the incoming and outgoing momenta of the heavy scalar of mass $M_\sigma$. The gravitational force will be mediated by massless gravitons of helicity two in a Lorentz frame where $\Delta^0=\Delta_z=0$. We will always work  to leading power in $M_\sigma$ and seek the first power corrections in $E_\phi$.
Our approximation will be expressed into power corrections of the form $\frac{\Delta}{E_\phi}=\frac{2\sqrt{-t}M_\sigma}{s-M_\sigma^2}$, where $s$, $t$ are the Mandelstam invariants. In our frame $p^0=p’^0=E_\phi=|\vec{p}|$ and $\vec{p}$ is nearly taken along the $z$-axis amounting to
\begin{equation}
p^\mu=\left(E_\phi, p^z, -\frac{\vec{\Delta}}{2}\right)\,\,\,\,\,\,\,\, p’^\mu=\left(E_\phi, p^z,\frac{\vec{\Delta}}{2}\right),
\label{quadri}
\end{equation}
with $p^z=\sqrt{E_\phi^2-\frac{\Delta^2}{4}}=p’^z$, because $p^z=E_\phi$ up to corrections of order $\frac{\Delta^2}{E_\phi^2}$, which may be neglected at the leading power of our approximation. 

\section{Path integral description by semiclassical gravity \label{DUE}}

The full four-point connected Green function of four scalars in quantum gravity is given by
\begin{eqnarray}
&&G(x_1,x’_1, x_2, x’_2)=\int\limits_{\substack{\text {Connected}\\ \text{diagrams}}} [Dh][D\phi_1] [D\phi_2] \phi_1(x_1)\phi_1(x’_1)\phi_2(x_2)\phi_2(x’_2)\nonumber\\
&&\exp\left\{i\int d^4 x \sqrt{-g}\left[\frac{1}{16\pi G_N}\left(R(h)-\frac{1}{2}g_{\mu\nu}C^\mu C^\nu\right)\right.\right.\nonumber\\&&
\left.\left.+\frac{1}{2}g^{\mu\nu}\partial_\mu\phi_1\partial_\nu\phi_1+\frac{1}{2}g^{\mu\nu}\partial_\mu\phi_2\partial_\nu\phi_2-\frac{1}{2}M_\sigma^2\phi_2^2\right]\right\},\nonumber\\
\label{amplitude}
\end{eqnarray}
where $R$ is the curvature scalar and $g_{\mu\nu}$ is defined as the sum of a flat Minkowski component $\eta_{\mu\nu}$ and a perturbation $\kappa h_{\mu\nu}$, with $\kappa= \sqrt{32\pi G_N}$ in terms of the Newton constant. It is coupled to the scalars matter represented by the fields $\phi_1$ and $\phi_2$, the first massless and the second heavy. $C_\mu$ fixes the gauge
\begin{equation}
C_\mu=\kappa\left(\partial_\nu h^{\nu}_\mu-\frac{1}{2}\partial_\mu h^\nu_\nu\right),
\label{gf}
\end{equation}
amounting to a covariant generalization of the de Donder gauge.
The ghosts for this gauge are not included because they do not contribute in the forward limit, $\Delta<< E_\phi$. The ghost-ghost-graviton vertex upon flat space, that can be found for instance in \cite{donnoghue}, is easily seen to be subdominant at every step of our analysis.
The path integral (\ref{amplitude}) can be rewritten as 
\begin{eqnarray}
&&G(x_1,x’_1, x_2, x’_2)=\int\limits_{\substack{\text {Connected}\\ \text{diagrams}}} [Dh] G^c(x_1,x’_1|h_{\mu\nu}) G^c_2(x_2,x’_2|h_{\mu\nu})\times\nonumber\\
&&\exp\left\{i\int d^4 x \sqrt{-g}\left[\frac{1}{16\pi G_N}\left(R(h)-\frac{1}{2}g_{\mu\nu}C^\mu C^\nu\right)\right]\right\},
\end{eqnarray}
where $G^c(x_1,x’_1|h_{\mu\nu})$ and $ G^c_2(x_2,x’_2|h_{\mu\nu})$ are two-point connected Green functions for the propagation of free scalar fields $\phi_1$ and $\phi_2$ of masses respectively $0$ and $M_\sigma$ in the presence of $h_{\mu\nu}$, taken as gravitational background field. In the eikonal approximation we will consider a gravitational interaction linearized in $h_{\mu\nu}$ and for the next-to-eikonal also the quadratic contributions in the background field, so called seagull terms, together with the trilinear gravitons vertex will be taken into account. The sum of the ladder diagrams, those not involving the self energy corrections, is based on the contributions to the scattering amplitude with the given kinematics (\ref{process}) of the on-shell two-point functions. For the light scalar
\begin{eqnarray}
&&<p’|G^c(x,y|h_{\mu\nu})|p>=\lim_{p^2,p’^2\rightarrow 0} \int d^4x  \int d^4 y\,\nonumber\\ &&e^{-ip\cdot x}\overrightarrow{(\partial_x^2)}(G(x,y|h)-G_0(x,y))\overleftarrow{(\partial_y^2)}e^{ip’\cdot y},
\label{connected}
\end{eqnarray}
where $G_0$ refers to the free propagation for the scalar field without any external background. Analogously for the heavy scalar of mass $M_\sigma$
\begin{eqnarray}
&&<q’|G^c(x,y|h_{\mu\nu})|q>=\lim_{q^2,q’^2\rightarrow M_\sigma^2} \int d^4x  \int d^4 y\, \nonumber\\&&e^{-iq\cdot x}\overrightarrow{(\partial_x^2)}(G(x,y|h)-G_0(x,y))\overleftarrow{(\partial_y^2)}e^{iq’\cdot y}.
\end{eqnarray}
The requested scattering amplitude leads to the $\mathcal{T}$ matrix element
\begin{eqnarray}
&&i(2\pi)^4\delta^4(p+q-p’-q’) \mathcal{T}(p,p’;q,q’)=\nonumber\\&&\exp\left [ \int d^4x d^4y \left(\frac{\delta}{\delta h^{\alpha\beta}(x)}D^{\alpha\beta,\gamma\delta}(x-y)\frac{\delta}{\delta h^{\prime \gamma\delta}(y)}\right)\right]\nonumber\\&&\left.<p’|G^c(x_1,x’_1|h)|p><q’|G^c(x_2,x’_2|h’)|q>\right |_{h,h’=0 }.
\label{T-matrix}
\end{eqnarray}
Here $D^{\mu\nu,\alpha\beta}$ is the graviton propagator so that
\begin{equation}
\int D_{\mu\nu,\alpha\beta}(x-z)D^{\alpha\beta,\gamma\delta}(z-y)d^4z=\frac{i}{2}\delta^4(x-y)(\eta_\mu\,^\gamma\eta_\nu\,^\delta +\eta_\mu\,^\delta\eta_\nu\,^\gamma )  \label{gravprop}
\end{equation}
which in the de Donder gauge specified by the gauge fixing terms in (\ref{amplitude}) and (\ref{gf}), takes the form
\begin{equation}
D_{\mu\nu,\alpha\beta}(x-y)=\int\frac{d^4 k}{(2\pi)^4}\frac{i}{2}\frac{\eta_{\mu \alpha}\eta_{\nu\beta}+\eta_{\mu \beta}\eta_{\nu\alpha}-\eta_{\mu\nu}\eta_{\alpha\beta}}{k^2}e^{-ik(x-y)}.
\label{propag}
\end{equation}
We will get $G^c(x,x’|h_{\mu\nu})$ in the next section by following the functional methods of Schwinger-Fradkin \cite{Fradkin1},\cite{Fradkin2}.

\section{Two point scalar connected Green function \label {TRE}}
We describe how to calculate the relativistic connected amputated Green function  $G^c(x,y|h_{\mu\nu})$ in a linearized gravitational background for a general scalar field of mass $m$.
The spin zero matter action is 
\begin{equation}
S=\int d^4 x\sqrt{-g}\left(\frac{1}{2}g^{\mu\nu}\partial_\mu\phi\partial_\nu\phi-\frac{1}{2}m^2\phi^2\right),
\label{matter}
\end{equation}
expanded untill trilinear couplings \cite{donnoghue} 
\begin{equation}
S=-\frac{1}{2}\int d^4 x \phi(x)\left[\partial^2 +m^2 -\kappa h^{\mu\nu}\partial_\mu\partial_\nu +\frac{\kappa}{2}\eta_{\mu\nu}h^{\mu\nu}\partial^2+\frac{\kappa}{2}\eta^{\mu\nu}h_{\mu\nu}m^2\right]\phi(x).
\end{equation}
The requested Green function is defined as
\begin{equation}
\left(\partial^2+m^2-\kappa h^{\mu\nu}\partial_\mu\partial_\nu +\frac{\kappa}{2}\eta_{\mu\nu}h^{\mu\nu}\partial^2 +m^2\frac{\kappa}{2}\eta_{\mu\nu}h^{\mu\nu}\right)G(x,y|h)=-i\delta^4(x-y).
\end{equation}
It is convenient to go over the momentum representation with respect to $x-y$ 
\begin{equation}
G(x,y|h)=\frac{1}{(2\pi)^4}\int d^4 \ell\, G(x,\ell)\,e^{-i\ell(x-y)},
\end{equation}
therefore 
\begin{eqnarray}
&&\left[-\ell^2+m^2+\partial^2 -2i\ell_\alpha\partial^\alpha +\kappa \ell_\mu \ell_\nu h^{\mu\nu}-\kappa h^{\mu\nu}\partial_\mu\partial_\nu +2i\kappa \ell_\mu h^{\mu\nu}\partial_\nu\right.\nonumber\\ &&\left.-\frac{\kappa}{2}\ell^2\eta_{\mu\nu}h^{\mu\nu}+\frac{\kappa}{2}\eta_{\mu\nu}h^{\mu\nu}\partial^2-i\kappa\eta_{\mu\nu}h^{\mu\nu}\ell_\alpha\partial^\alpha+\frac{\kappa}{2}\eta_{\mu\nu}h^{\mu\nu}m^2\right]G(x,\ell)=-i.\nonumber\\
\end{eqnarray}
In terms of the Schwinger’s proper time $\nu$ as well as the Feynman $i\epsilon$ prescription for the propagator 
\begin{equation}
G(x,\ell)=\int_0^{+\infty} e^{i\nu(\ell^2-m^2+i\epsilon)}Y(x,\nu)d\nu.
\end{equation}
We have the following Cauchy’s problem
\begin{equation}
\begin{cases} 
i\frac{\partial Y}{\partial\nu}(x,\nu)=\left[\partial^2-2i \ell_\mu\partial^\mu +\kappa \ell_\mu \ell_\beta h^{\mu\beta}-\kappa h^{\mu\beta}\partial_\mu\partial_\beta +2i\kappa \ell_\mu h^{\mu\beta}\partial_\beta\right.\\
\left.-\frac{\kappa}{2}\ell^2\eta_{\mu\beta}h^{\mu\beta}+\frac{\kappa}{2}\eta_{\mu\beta}h^{\mu\beta}\partial^2 -i\kappa \eta_{\mu\beta}h^{\mu\beta}\ell_\alpha\partial^\alpha+\frac{\kappa}{2}\eta_{\mu\beta}h^{\mu\beta}m^2\right]Y(x,\nu) \vspace*{0.2cm} \\ 
Y(x,0)=1,
\end{cases}
\end{equation}
where the Feynman prescription for the propagator has been included. Take the external gravitational field $h_{\alpha\beta}(x)=\hat{h}_{\alpha\beta}(k,\frac{\partial}{\partial k})e^{ikx}$, where $\hat{h}_{\alpha\beta}(k,\frac{\partial}{\partial k})$ is an arbitrary operator of $k$ and derivatives with respect to $k$ with the only restriction that $k^2=0$ and $k^\alpha \hat{h}_{\alpha\beta}=\frac{1}{2}k_\beta\hat{h}^\rho\,_\rho(k,\frac{\partial}{\partial k})$. By taking into account that $\partial^2\approx k^2=0$ and also $h^\alpha\,_\beta\partial_\alpha\approx ik^\alpha \hat{h}_{\alpha\beta} e^{ikx}=i \frac{1}{2}k_\beta\hat{h}^\rho\,_\rho(k,\frac{\partial}{\partial k})e^{ikx}$. The above approximations are equivalent to state that the spatial variation of $Y(x,\nu)$ is approximately driven by the gravitational plane wave background like in a forced harmonic oscillator \cite{georgi}. The Cauchy’s problem changes as
\begin{equation}
\begin{cases} 
i\frac{\partial Y}{\partial\nu}(x,\nu)=\left[\partial^2-2i \ell_\alpha\partial^\alpha +\kappa \ell_\mu \ell_\beta h^{\mu\beta}(x)\right]Y(x,\nu) \vspace*{0.2cm} \\ 
Y(x,0)=1,
\end{cases}
\label{Cauchy1}
\end{equation}
where the on-shell limit, $\ell^2=m^2$, has been used since we will need that limit to compute the $\mathcal{T}$-matrix element of (\ref{T-matrix}).
Let’s introduce an additional interaction with the external generators  $t_\mu (\nu)$ of the operators $\partial_\mu$: 
\begin{equation}
i\frac{\partial Y}{\partial\nu}(x,\nu,t)=\left[\partial^2 -2i \ell_\alpha\partial^\alpha +\kappa \ell_\mu \ell_\beta h^{\mu\beta}(x)+ it_\mu(\nu)\partial^\mu\right]Y(x,\nu,t)
\label{Schroe}
\end{equation}
and
\begin{equation}
\left.\frac{\delta}{\delta t^\mu (\nu)}Y\right|_{t=0}= \partial_\mu Y(\nu).
\end{equation}
Treating the Schrödinger type equation (\ref{Schroe}) by functional methods \cite{peskin},\cite{mazzucchi}
\begin{equation}
\left.Y(x,\nu)=\exp\left[-i\int_0^\nu \frac{\delta^2}{\delta t^\mu (\xi)\delta t_\mu (\xi)}d\xi\right] Y_1(x,\nu,t)\right|_{t=0}
\end{equation}
where
\begin{equation}
\frac{\partial Y_1}{\partial \nu}(x,\nu,t)= (-2\ell_\alpha\partial^\alpha -i\kappa \ell_\mu \ell_\beta h^{\mu\beta}(x)+t_\mu(\nu)\partial^\mu)Y_1(x,\nu,t).
\end{equation}
By the method of variation of arbitrary constants
\begin{equation}
Y_1(x,\nu,t)=\exp\left(-\int_0^\nu (2\ell_\alpha\partial^\alpha-t_\mu(\xi)\partial^\mu)d\xi\right)Y_2(x,\nu,t),
\label{Y1}
\end{equation}
with
\begin{equation}
\frac{\partial Y_2}{\partial \nu}=-i\kappa \ell_\mu \ell_\beta h^{\mu\beta}\left(x+2\ell\nu-\int_0^\nu t(\xi)d\xi\right)Y_2,
\end{equation}
giving therefore
\begin{equation}
Y_2=\exp\left[-i\kappa \ell_\mu \ell_\beta \int_0^\nu h^{\mu\beta}\left(x+2\ell\nu’-\int_0^{\nu’} t(\xi)d\xi\right)d\nu’\right].
\end{equation}
By taking into account the translation operator of (\ref{Y1})
\begin{eqnarray}
&&Y(x,\nu)= \exp\left[-i\int_0^\nu  \frac{\delta^2}{\delta t^\mu (\xi)\delta t_\mu (\xi)}d\xi\right]\nonumber\\&&\left.\exp\left[-i\kappa \ell_\mu \ell_\beta 
\int_0^\nu h^{\mu\beta}\left(x+2\ell(\nu’-\nu)+\int_{\nu’}^{\nu} t(\xi)d\xi\right)d\nu’\right]\right|_{t=0},
\end{eqnarray}
amounting to
\begin{eqnarray}
&& Y(x,\nu)= \exp\left[-i\int_0^\nu  \frac{\delta^2}{\delta t^\mu (\xi)\delta t_\mu (\xi)}d\xi\right]\times \nonumber \qquad \qquad \\ &&  \qquad \qquad \, \, \, \exp\left[-i\kappa \ell_\mu \ell_\beta \int_0^\nu h^{\mu\beta}\left(x+2\ell(\nu’-\nu) \right. \right. \nonumber \\ && \left. \left. \left. \qquad \qquad \qquad \quad \qquad +  \quad \int_{0}^{\nu} t(\xi)\theta (\xi-\nu’)d\xi\right)d\nu’\right]\right|_{t=0}.
\end{eqnarray} 
In terms of the Fourier modes of the gravitational field 
\begin{align}
Y(x,\nu)= & \sum_{n=0}^{+\infty}\frac{(-i\kappa)^{n}}{n!}\prod_{i,j=1}^{n}\int_{0}^{\nu}d\xi_{i}\int\frac{d^{4}k_{j}}{(2\pi)^{4}}\ell_{\mu_{i}}\ell_{\beta_{i}}\hat{h}^{\mu_{j}\beta_{j}}(k_{j})\times\nonumber \\
 & \,\exp\left(-ik_{j}\cdot x-2i\ell\cdot k_{j}(\xi_{i}-\nu)\right)\times\label{eq:Yxnu}\\
 & \left.\exp\left[-i\int_{0}^{\nu}\frac{\delta^{2}}{\delta t^{\mu}(\xi)\delta t^{\mu}(\xi)}d\xi\right]\exp\left[-i\int_{0}^{\nu}t(\xi)\cdot k_{j}\theta(\xi-\nu\text{\textquoteright})d\xi\right]\right|_{t=0}\nonumber 
\end{align}
The following property holds 
\begin{align}
 & \exp\left[i\int_{0}^{\nu}g^{\rho}(\xi)g_{\rho}(\xi)d\xi\right]= \nonumber \\
 & \left.\exp\left[-i\int_{0}^{\nu}\frac{\delta^{2}}{\delta t^{\mu}(\xi)\delta t_{\mu}(\xi)}d\xi\right]\exp\left[i\int_{0}^{\nu}t^{\mu}(\xi)g_{\mu}(\xi)d\xi\right]\right|_{t=0} \label{PROP1} 
\end{align} 
which is proved in \cite{Fried} and \cite{sommerfeld}. 
Finally 
\begin{align}
Y(x,\nu)= & 1+\sum_{n=1}^{+\infty}\frac{(-i)^{n}\kappa^{n}}{n!}\left(\prod\limits _{m=1}^{n}\int_{0}^{\nu}d\xi_{m}\int\frac{d^{4}k_{m}}{(2\pi)^{4}}\ell^{\mu}\ell^{\beta}\hat{h}_{\mu\beta}(k_{m})e^{-ik_{m}\cdot x}\right)\times\nonumber \\
 & \exp\left[i\sum_{m,m_{1}}k_{m}\cdot k_{m_{1}}\left(\frac{\xi_{m}+\xi_{m_{1}}}{2}+\frac{1}{2}|\xi_{m}-\xi_{m_{1}}|-\nu\right)\right]\times\nonumber \\
 & \exp\left[2i\sum_{m=1}^{n}\ell\cdot k_{m}(\nu-\xi_{m})\right],\label{amplY}
\end{align}
which is readily obtained by using (\ref{PROP1}) together with 
\begin{equation}
\int_0^\nu \theta(\xi-\xi_m)\theta(\xi-\xi_{m_1})d\xi=\nu-\frac{\xi_m+\xi_{m_1}}{2}-\frac{|\xi_m-\xi_{m_1}|}{2}.
\end{equation} 
\section{Eikonal approximation\label{QUATTRO}}

In the forward limit, $p\sim p’$ the amplitude (\ref{connected}) gets the following expression 
\begin{align}
<p\text{\textquoteright}|G^{c}(x,y|h)|p> & =\lim\limits _{\substack{p^{2}\rightarrow0\\
p^{\prime^{2}}\rightarrow0
}
}\int d^{4}xd^{4}ye^{-ip\cdot x}\overrightarrow{\partial_{x}^{2}}\int\frac{d^{4}\ell}{(2\pi)^{4}}e^{-i\ell\cdot(x-y)}\times\nonumber \\
 & \qquad\qquad\int_{0}^{+\infty}d\nu e^{i\nu(\ell^{2}+i\epsilon)}(Y(x,\nu)-1)\overleftarrow{\partial_{y}^{2}}e^{ip\text{\textquoteright}\cdot y}.\label{Gc}
\end{align}
In the eikonal limit where the terms $k^2$ and $k_i\cdot k_j$ type are neglected, more precisely $|k_i\cdot k_j|<<|p\cdot k_i|$ or $|p\cdot k_j|$, we have
\begin{align}
 & <p\text{\textquoteright}|G^{c}(x,y|h)|p>=\nonumber \\
 & \lim\limits _{\substack{p^{2}\rightarrow0\\
p^{\prime2}\rightarrow0
}
}\int d^{4}xd^{4}ye^{-ip\cdot x}\overrightarrow{\partial_{x}^{2}}\int\frac{d^{4}\ell}{(2\pi)^{4}}e^{-i\ell\cdot(x-y)}\int_{0}^{+\infty}d\nu e^{i\nu(\ell^{2}+i\epsilon)}\times\nonumber \\
 & \qquad\sum_{n=1}^{+\infty}\frac{(-i)^{n}\kappa^{n}}{n!}\prod\limits _{m=1}^{n}\int_{0}^{\nu}d\xi_{m}\int\frac{d^{4}k_{m}}{(2\pi)^{4}}\ell^{\mu_{m}}\ell^{\beta_{m}}\hat{h}_{\mu_{m}\beta_{m}}(k_{m})\times\nonumber \\
 & \qquad e^{-ik_{m}\cdot x}e^{2i\sum\limits _{m=1}^{n}\ell\cdot k_{m}(\nu-\xi_{m})}\overleftarrow{\partial_{y}^{2}}e^{ip\text{\textquoteright}\cdot y}\quad = \nonumber \\
 & \lim\limits _{\substack{p^{2}\rightarrow0\\
p^{\prime2}\rightarrow0
}
}p^{2}p^{\prime^{2}}(2\pi)^{4}\sum_{n=1}^{+\infty}\frac{(-i)^{n}\kappa^{n}}{n!}\int_{0}^{+\infty}d\nu e^{i\nu(p^{\text{\textquoteright}2}+i\epsilon)}\times\nonumber \\
 & \qquad\int\frac{d^{4}k_{1}\dots d^{4}k_{n}}{(2\pi)^{4n}}\int_{0}^{\nu}d\xi_{1}\dots d\xi_{n}\delta^{4}(p\text{\textquoteright}-p+k_{1}+\dots k_{n})\times\nonumber \\
 & \qquad e^{-\sum\limits _{m=1}^{n}2ip^{\prime}\cdot k_{m}(\nu-\xi_{m})}p^{\prime\mu_{1}}p^{\prime\beta_{1}}\hat{h}_{\mu_{1}\beta_{1}}(k_{1})\dots p^{\prime\mu_{n}}p^{\prime\beta_{n}}\hat{h}_{\mu_{n}\beta_{n}}(\tilde{k}_{r}).\label{eikonallimitGc}
\end{align}
For the heavy line we have 
\begin{eqnarray}
&&  \lim\limits_{\substack {q^2\rightarrow M_\sigma^2\\ q^{\prime^2}\rightarrow M_\sigma^2}}(q^2-M_\sigma^2)(q^{\prime^2}-M_\sigma^2)(2\pi)^4\sum_{r=1}^{+\infty}\frac{(-i)^r\kappa^r}{r!}\int_0^{+\infty}d\nu_1 e^{i\nu_1(q’^2-M_\sigma^2+i\epsilon)}\nonumber\\&&\int\frac{d^4 \tilde{k}_1\dots d^4\tilde{k}_r}{(2\pi)^{4r}}\int_0^{\nu_1} d\xi_1\dots\nonumber\int_0^{\nu_1} d\xi_r \delta^4(q’-q+\tilde{k}_1+\dots \tilde{k}_r)\times \nonumber\\&& e^{-\sum\limits_{m=1}^r 2i q’\cdot \tilde{k}_m(\nu_1-\xi_m)} q’^{\mu_1}q’^{\beta_1}\hat{h}_{\mu_1\beta_1}(\tilde{k}_1)\dots q’^{\mu_r}q’^{\beta_r}\hat{h}_{\mu_r\beta_r}(\tilde{k}_r). \label{HL}
\end{eqnarray}
The integral on a $\xi_m$ variable is
\begin{equation}
\int_0^\nu d\xi_m \exp[-2ip’\cdot k_m(\nu-\xi_m)]=\frac{1}{2ip’\cdot k_m}\left[1-\exp(-2i\nu p’\cdot k_m)\right].
\end{equation}
The eikonal  $i\mathcal{T}$ matrix element for the process (\ref{process}) in the specified kinematics is obtained from (\ref{T-matrix}) 
\begin{align}
 & i\mathcal{T}(p,p\text{\textquoteright},q,q\text{\textquoteright})=\lim\limits _{p_{i}^{2}\rightarrow m_{i}^{2}}p^{2}p\text{\textquoteright}^{2}(q^{2}-M_{\sigma}^{2})(q\text{\textquoteright}^{2}-M_{\sigma}^{2})(2\pi)^{4}\times\nonumber \\
 & \sum_{n=1}^{+\infty}\frac{(-1)^{n}\kappa^{2n}}{n!}\int_{0}^{+\infty}d\nu\,e^{i\nu(p\text{\textquoteright}^{2}+i\epsilon)}\int_{0}^{+\infty}d\nu_{1}\,e^{i\nu_{1}[(q\text{\textquoteright}^{2}-M_{\sigma}^{2})+i\epsilon]}\nonumber \\
 & \prod_{m=1}^{n}\frac{i}{2}\int\frac{d^{4}k_{m}}{(2\pi)^{4}}p^{\prime\mu_{m}}p^{\prime\beta_{m}}\frac{2\eta_{\mu_{m}0}\eta_{\beta_{m}0}-\eta_{\mu_{m}\beta_{m}}\eta_{00}}{k_{m}^{2}}\frac{1}{2ip\text{\textquoteright}\cdot k_{m}}M_{\sigma}^{2}\frac{1}{-2iq\text{\textquoteright}\cdot k_{m}}\nonumber \\
 & \left[1-e^{(-2i\nu p\text{\textquoteright}\cdot k_{m})}\right]\left[1-e^{(2i\nu_{1}q\text{\textquoteright}\cdot k_{m})}\right]\delta^{4}(q\text{\textquoteright}-q-k_{1}-k_{2}-\dots k_{n}),\label{T}
\end{align} 
where by $\lim\limits_{p_i^2\rightarrow m_i^2} $ we synthetically refer to all on the mass shell limits involved in (\ref{T}). The expression of the momentum space propagator of gravitons is in~(\ref{propag}).
We are going to use the so called eikonal identity \cite{Levy},\cite{cardy},
\begin{eqnarray}
&&\sum_\pi \frac{1}{x+i\epsilon}\frac{1}{x+a_{\pi(1)}+i\epsilon}\dots \frac{1}{x+a_{\pi(1)}+\dots a_{\pi(n)}+i\epsilon}=\nonumber\\&& (-i)\int_0^{+\infty}d\nu e^{i\nu x}\prod_{m} \frac{1-e^{i\nu a_m}}{a_m+i\epsilon},
\label{eik}
\end{eqnarray}
where $\pi$ belongs to the set of permutations of $n$ indices. The sum of the permutations is done by the use of the identity proved in \cite{Sterman},
\begin{eqnarray}
&&\sum_{perms\, of\, \omega_i} \delta(\omega_1+\dots\omega_n)\frac{1}{\omega_1+i\epsilon}\dots \frac{1}{\omega_1+\dots +\omega_{n-1}+i\epsilon}\nonumber\\&&
=\delta(\omega_1+\dots\omega_n)\sum_{\omega_n}\prod_{i=1}^{n-1}\frac{1}{\omega_j+i\epsilon}=(-2\pi i)^{n-1}\delta(\omega_1)\dots\delta(\omega_n).
\label{eik1}
\end{eqnarray}
To see how the previous identities work together with the on-shell limit, consider the heavy scalar line and take 
\begin{eqnarray}
&&-i\int_0^{+\infty} d\nu_1 e^{i\nu_1[q^{\prime 2}-M_\sigma^2+i\epsilon]}\prod_{m=1}^n \frac{1-\exp(2i\nu_1q’\cdot k_m)}{2q’\cdot k_m}= \sum_{\pi}\frac{1}{q^{\prime 2} -M_\sigma^2+i\epsilon}\nonumber\\&&\frac{1}{q^{\prime 2} -M_\sigma^2+2q’\cdot k_{\pi(1)}+i\epsilon}\dots \frac{1}{q^{\prime 2} -M_\sigma^2+2q’\cdot (k_{\pi(1)}+\dots k_{\pi(n)})+i\epsilon}.\nonumber\\
\label{eik2}
\end{eqnarray}
By using the delta function of momentum conservation $\delta^4(q’-q-k_1-k_2-\dots k_n)$ the last fraction in the static limit, $q\sim q’ = (M_\sigma,0)$, amounts to
\begin{equation}
 \frac{1}{q^{\prime 2} -M_\sigma^2+2q’\cdot (q’-q)}\sim \frac{1}{q^2-M_\sigma^2},
\label{eik22}
\end{equation}
consequently
\begin{eqnarray} 
&&\lim\limits_{\substack {q^2\rightarrow M_\sigma^2\\ q’^2\rightarrow M_\sigma^2}}\delta(k^0_1+\dots k^0_n)(q^2-M_\sigma^2)(q^{\prime 2}-M_\sigma^2)\nonumber\\&& \sum_{\pi}\frac{1}{q^{\prime 2} -M_\sigma^2+i\epsilon}\dots  \frac{1}{q^{\prime 2} -M_\sigma^2+2q’\cdot (k_{\pi(1)}+\dots k_{\pi(n)})}=\nonumber\\
&&\frac{(-2\pi i)^{n-1}}{(2M_\sigma)^{n-1}}\delta(k^0_1)\dots\delta(k^0_n).
\label{eik3}
\end{eqnarray}
By the use of (\ref{eik}), (\ref{eik1}), (\ref{eik2}), (\ref{eik22}) and (\ref{eik3}) the amplitude (\ref{T}) becomes 
\begin{align}
 & i\mathcal{T}(p,p\text{\textquoteright};q,q\text{\textquoteright})=\nonumber \\
 & \lim\limits _{p_{i}^{2}\rightarrow m_{i}^{2}}p^{2}p\text{\textquoteright}^{2}(2\pi)^{4}\sum_{n=1}^{+\infty}\frac{(-1)^{n}\kappa^{2n}}{n!}\int_{0}^{+\infty}d\nu\,e^{i\nu(p\text{\textquoteright}^{2}+i\epsilon)}\delta^{3}(\vec{\Delta}+\vec{k}_{1}+\vec{k}_{2}+\dots\vec{k}_{n})\nonumber \\
 & \left(\prod_{m=1}^{n}\int\frac{d^{3}k_{m}}{(2\pi)^{4}}\frac{i}{\vec{k}_{m}^{2}}E_{\phi}^{2}M_{\sigma}^{2}\frac{1-\exp(2i\nu\vec{p}^{\prime}\cdot\vec{k}_{m})}{2i\vec{p}^{\prime}\cdot\vec{k}_{m}}\right)\frac{(-2\pi i)^{n-1}}{(2M_{\sigma})^{n-1}}.\label{T1}
\end{align}
 Being the components of three-vector $\vec{k}_m=(k^z_m, \vec{k}^\perp_m)$, in our reference frame, $\Delta^z=0$, we can consider the expansion at the first order in $\frac{\vec{\Delta}\cdot \vec{k}_m^{\perp}}{2E_\phi k_m^z}$ for
\begin{equation}
\prod_{m=1}^n  \frac{1-\exp(2i\nu \vec{p}\,\,’\cdot \vec{k}_m)}{2i\vec{p}\,’\cdot \vec{k}_m}.
\end{equation}
From (\ref{quadri})
for our desired approximation we need
\begin{equation}
2i \vec{p}\,\,’\cdot \vec{k}_m = 2i E_\phi k_m^z +i\vec{\Delta}\cdot \vec{k}_m^\perp.
\end{equation}
Therefore up to the linear order in $\vec{\Delta}$
\begin{eqnarray}
&&\prod_{m=1}^n  \frac{1-\exp(2i\nu \vec{p}\,\,’\cdot \vec{k}_m)}{2i\vec{p}\,’\cdot \vec{k}_m}=\prod_{m=1}^n \frac{1-e^{2i\nu E_\phi k_m^z}}{2i E_\phi k_m^z}+\nonumber\\&&
\sum_{m=1}^n i\nu \vec{\Delta}\cdot \vec{k}_m^\perp \prod_{l=1}^n \frac{1-e^{2i\nu E_\phi k_l^z}}{2i E_\phi k_l^z} -\nu \sum_{m=1}^n\frac{\vec{\Delta}\cdot \vec{k}_m^\perp}{2E_\phi k_m^z}\prod_{l=1, l\neq m}^n \frac{1-e^{2i\nu E_\phi k_l^z}}{2i E_\phi k_l^z}\nonumber\\
&& -\sum_{m=1}^n \frac{\vec{\Delta}\cdot \vec{k}_m^\perp}{2k_m^z E_\phi} \prod_{l=1}^n \frac{1-e^{2i\nu E_\phi k_l^z}}{2i E_\phi k_l^z}.
\label{exp1}
\end{eqnarray}
In (\ref{exp1}) the term  proportional to $\sum\limits_{m=1}^n \vec{\Delta}\cdot \vec{k}_m^\perp$ is of order ${\Delta}^2$ due to the $\delta^2(\vec{\Delta}+\vec{k}_1+\vec{k}_2+\dots \vec{k}_n)$ and therefore it is disregarded in our approximation. The last two terms in the l.h.s of (\ref{exp1}) give a vanishing contribution to the scattering amplitude since they cancel each others, because those terms are dominated by $k_m^z\rightarrow 0$ 
\begin{equation}
\lim_{k_m^z\rightarrow 0} \frac{1-e^{2i\nu E_\phi k_m^z}}{2i E_\phi k_m^z}=-\nu,
\end{equation}
consequently (\ref{exp1}) is approximated as
\begin{equation}
\prod_{m=1}^n  \frac{1-\exp(2i\nu \vec{p}\,\,’\cdot \vec{k}_m)}{2i\vec{p}\,’\cdot \vec{k}_m}\sim\prod_{m=1}^n \frac{1-e^{2i\nu E_\phi k_m^z}}{2i E_\phi k_m^z}
\end{equation}
from which the amplitude in the eikonal approximation is obtained by applying the eikonal identities (\ref{eik}), (\ref{eik1}) with the total null momentum condition along the $z$-axis $\delta(k_1^z+\dots +k_n^z)$: 
\begin{align}
 & i\mathcal{T}(p,p\text{\textquoteright};q,q\text{\textquoteright})\equiv i\mathcal{T}(\vec{\Delta})=\label{amplT}\\
 & -4(2\pi)^{2}iE_{\phi}M_{\sigma}\sum_{n=1}^{+\infty}\frac{1}{n!}\left(\frac{i\kappa^{2}E_{\phi}M_{\sigma}}{16\pi^{2}}\right)^{n}\prod_{m=1}^{n}\int\frac{d^{2}k_{m}^{\perp}}{\vec{k}_{m}^{\perp2}}(2\pi)^{2}\delta^{2}(\vec{k}_{1}^{\perp}+\dots\vec{k}_{n}^{\perp}+\vec{\Delta}).\nonumber 
\end{align}
The expression found in (\ref{amplT}) is dimensionless in natural units as it must be for a $2\rightarrow 2$ process. Now let us Fourier transform into the transverse impact parameter space,
\begin{eqnarray}
&&i\widetilde{ \mathcal{T}}(\vec{b}^{\perp})\equiv \int \frac{d^2\Delta}{(2\pi)^2}e^{i\vec{b}^{\perp}\cdot \vec{\Delta}}\,i \mathcal{T}(\Delta)=\nonumber\\&& 
-4i E_\phi M_\sigma\sum_{n=1}^{+\infty}\frac{1}{n!}\left(\frac{i\kappa^2 E_\phi M_\sigma}{16\pi^2}\right)^n\int d^2k_1^{\perp}\dots 
d^2k_n^{\perp}\prod_{i=1}^n\left[\frac{e^{-i\vec{b}^{\perp}\cdot \vec{k}_i^{\perp}}}{\vec{k}_i^{\perp 2}}\right],
\label{impact}
\end{eqnarray} 
which exponentiates as
 \begin{equation}
i\widetilde{ \mathcal{T}}(\vec{b}^{\perp})= -4iE_\phi M_\sigma (e^{i\chi_0}-1)
\end{equation}
being the eikonal phase
\begin{equation}
\chi_0 ({\vec{b}}^\perp)= \frac{\kappa^2 M_\sigma E_\phi}{16\pi^2}\int \frac{d^2 k^{\perp}}{\vec{k}^{\perp 2}}e^{-i\vec{b}^{\perp}\cdot \vec{k}^{\perp}}.
\label{fase}
\end{equation}
In dimensional regularization by continuing to $d>2$ dimensions \cite{Strichartz}
\begin{equation}
\int d^d k \frac{e^{-i\vec{b}\cdot\vec{k}}}{\vec{k}^2}=\pi^{\frac{d}{2}}2^{-2+\frac{d}{2}}\Gamma\left(-1+\frac{d}{2}\right)|\vec{b}|^{2-d}.
\end{equation} 
The eikonal phase is indeed an infrared dominated quantity and a finite result for the scattering amplitude is obtained after the resummation \cite{Giddings}.
In the impact parameter space (\ref{impact}) 
is dominated by the solution of the saddle point equation for the integral on $\vec{b}^\perp$
\begin{equation}
|\vec{\Delta}|\cos\theta - 2G M_\sigma E_\phi |\vec{b}^\perp|^{1-d}=0
\end{equation}
and at $d=2$ the order of magnitud of the saddle point is
\begin{equation}
|\vec{b}^{\perp}|\sim \frac{GM_\sigma E_\phi}{|\vec{\Delta}|}.
\label{bigimpact}
\end{equation}
In the ultrarelativistic limit at small momentum transfer we have a large impact parameter with respect to the Schwarzschild radius of the target particle, $R_s= GM_\sigma$, therefore the contributions around $\vec{b}=0$ will be disregarded. We are now going to explore this regime of scattering of large impact parameter by removing the eikonal approximation. 

\section{Next-to-eikonal approximation\label{CINQUE}}
From (\ref{amplY}) the next to eikonal corrections to the scattering amplitude in which the leading quadratic dependence upon the gravitons virtual momenta is taken, receives the following contributions
\begin{eqnarray}
&&i\mathcal{T}(p,p’,q,q’)_{NE}\equiv\lim_{p_i^2\rightarrow m_i^2} p^2p’^2(q^2-M_\sigma^2)(q’^2-M_\sigma^2)\sum_{n=1}^{+\infty}\frac{(-1)^n\kappa^{2n}}{n!}\nonumber\\&&\int_0^{+\infty}d\nu e^{i\nu(p’^2+i\epsilon)}\int_0^{+\infty}d\nu_1 e^{i\nu_1[(q’^2-M_\sigma^2)+i\epsilon]}\prod_{m=1}^{n}\int\frac{d^4 k_m}{(2\pi)^4} E_\phi^2 M_\sigma^2\frac{i}{k_m^2}\nonumber\\&&\int_0^{\nu}d\xi_1\dots \int_0^{\nu}d\xi_n \int_0^{\nu_1}d\tilde{\xi}_1\dots \int_0^{\nu_1}d\tilde{\xi}_n (2\pi)^4\delta^4(q-q’-k_1-\dots k_n)\nonumber\\&& (-i)\exp\left[\sum_{m=1}^n (-2ip’)\cdot k_m(\nu-\xi_m)\right]\exp\left[\sum_{\tilde{m}=1}^n (2iq’)\cdot k_{\tilde{m}}(\nu_1-\tilde{\xi}_{\tilde{m}})\right]\nonumber\\&&\left[\sum_{m, m_1=1}^n k_m\cdot k_{m_1}\left(\nu-\frac{\xi_m+\xi_{m_1}}{2}-\frac{|\xi_m-\xi_{m_1}|}{2} \right)\right.\nonumber\\&&\left.+\sum_{m_2, m_3=1}^n k_{m_2}\cdot k_{m_3}\left(\nu_1-\frac{\tilde{\xi}_{m_2}+\tilde{\xi}_{m_3}}{2}-\frac{|\tilde{\xi}_{m_2}-\tilde{\xi}_{m_3}|}{2} \right) \right].
\label{NE}
\end{eqnarray}
Observe that 
\begin{eqnarray}
&&\sum_{m, m_1=1}^n k_m\cdot k_{m_1}\left(\nu-\frac{\xi_m+\xi_{m_1}}{2}-\frac{|\xi_m-\xi_{m_1}|}{2} \right)\nonumber\\&&+\sum_{m_2, m_3=1}^n k_{m_2}\cdot k_{m_3}\left(\nu_1-\frac{\tilde{\xi}_{m_2}+\tilde{\xi}_{m_3}}{2}-\frac{|\tilde{\xi}_{m_2}-\tilde{\xi}_{m_3}|}{2} \right)=\nonumber\\&& \sum_{m=1}^n k_m^2 (\nu-\xi_m)+\sum_{m_1=1}^n k_{m_1}^2 (\nu_1-\tilde{\xi}_{m_1})\nonumber\\&& +\sum_{\substack{m_2, m_3=1\\m_2\neq m_3}}^n k_{m_2}\cdot k_{m_3}\left(\nu-\frac{\xi_{m_2}+\xi_{m_3}}{2}-\frac{|\xi_{m_2}-\xi_{m_3}|}{2} \right)\nonumber\\&& + \sum_{\substack{m_4, m_5=1\\m_4\neq m_5}}^n k_{m_4}\cdot k_{m_5}\left(\nu_1-\frac{\tilde{\xi}_{m_4}+\tilde{\xi}_{m_5}}{2}-\frac{|\tilde{\xi}_{m_4}-\tilde{\xi}_{m_5}|}{2} \right).
 \end{eqnarray}
Firstly consider the following contribution in dimensional regularization
\begin{eqnarray}
&&i\mathcal{T}(p,p’,q,q’)^1_{NE}\equiv\lim_{p_i^2\rightarrow m_i^2} p^2p’^2(q^2-M_\sigma^2)(q’^2-M_\sigma^2)\sum_{n=1}^{+\infty}\frac{(-1)^n\kappa^{2n}}{n!}\nonumber\\&&\int_0^{+\infty}d\nu e^{i\nu(p’^2+i\epsilon)}\int_0^{+\infty}d\nu_1 e^{i\nu_1[(q’^2-M_\sigma^2)+i\epsilon]}\prod_{m=1}^{n}\int\frac{d^{4-2\epsilon} k_m}{(2\pi)^{4}}\nonumber\\&& p^{\prime\mu_m}p^{\prime\beta_m}\frac{i}{2}\frac{2\eta_{\mu_m 0}\eta_{\beta_m 0}-\eta_{\mu_m\beta_m}\eta_{00}}{k_m^2}M_\sigma^2\int_0^{\nu}d\xi_1\dots \int_0^{\nu}d\xi_n \int_0^{\nu_1}d\tilde{\xi}_1\dots\nonumber\\&& \int_0^{\nu_1}d\tilde{\xi}_n (-i)\exp\left[\sum_{m=1}^n (-2ip’)\cdot k_m(\nu-\xi_m)\right]\exp\left[\sum_{\tilde{m}=1}^n (2iq’)\cdot k_{\tilde{m}}(\nu_1-\tilde{\xi}_{\tilde{m}})\right]\nonumber\\&&\sum_{r=1}^n k_r^2 (\nu-\xi_r)(2\pi)^{4}\delta^{4}(q-q’-k_1-\dots k_n). \label{NE1}
\end{eqnarray}
By using the formulas (\ref{eik1}), (\ref{eik2}), (\ref{eik22}) we reach
\begin{eqnarray}
&&\lim_{p_i^2\rightarrow m_i^2} p^2p’^2\sum_{n=1}^{+\infty}\frac{(-1)^n\kappa^{2n}}{n!}\int_0^{+\infty}d\nu e^{i\nu(p’^2+i\epsilon)}\prod_{m=1}^{n}\int\frac{d^{3-2\epsilon} k_m}{(2\pi)^{4}} (p^{\prime 0} M_\sigma)^2\frac{-i}{\vec{k}_m^2}\nonumber\\&&\int_0^{\nu}d\xi_1\dots \int_0^{\nu}d\xi_n (+i)\exp\left[\sum_{s=1}^n 2i\vec{p^{\prime}}\cdot \vec{k}_s(\nu-\xi_s)\right]\sum_{r=1}^n \vec{k}_r^2 (\nu-\xi_r)\frac{(-2\pi i)^{n-1}}{(2M_\sigma)^{n-1}}\nonumber\\&&(2\pi)^{4}\delta^{3}(\vec{\Delta}+\vec{k}_1+\dots \vec{k}_n). \label{NE2}
\end{eqnarray}
The following identity holds \cite{Levy},\cite{peskin},\cite{cardy},
\begin{eqnarray}
&&\int_0^\nu d\xi_1\dots\int_0^\nu d\xi_n \sum_{r=1}^n \vec{k}_r^2 (\nu-\xi_r)\exp\left[\sum_{s=1}^n 2i\vec{p^{\prime}}\cdot \vec{k}_s(\nu-\xi_s)\right]=\nonumber\\&&
\sum_{r=1}^n  \vec{k}_r^2 \frac{\partial}{\partial (2i\vec{p^{\prime}}\cdot \vec{k}_r)}\prod_{s=1}^n \frac{-1}{2i\vec{p^{\prime}}\cdot \vec{k}_s}[1-e^{2i\nu\vec{p^\prime}\cdot \vec{k}_s}]. 
\end{eqnarray}
By consequence
\begin{eqnarray}
&&\lim_{p_i^2\rightarrow m_i^2} p^2p’^2\int_0^{+\infty} d\nu e^{i\nu(p’^2+i\epsilon)}\prod_{s=1}^n \frac{-1}{2i\vec{p^{\prime}}\cdot \vec{k}_s}[1-e^{2i\nu\vec{p^\prime}\cdot \vec{k}_s}]=\nonumber\\
&& i^n \frac{1}{\vec{p^\prime}\cdot\vec{k}_1} \frac{1}{\vec{p^\prime}\cdot\vec{k}_2}\dots  \frac{1}{\vec{p^\prime}\cdot\vec{k}_{n-1}}
\end{eqnarray}
where the eikonal identities (\ref{eik}) and (\ref{eik1}) have been used. The contribution to the amplitude is therefore 
\begin{align}
 & \sum_{n=1}^{+\infty}\frac{(-1)^{n}\kappa^{2n}}{n!}\frac{i(-2\pi i)^{n-1}}{(2M_{\sigma})^{n-1}}\prod_{m=1}^{n}\int\frac{d^{3-2\epsilon}k_{m}}{(2\pi)^{4}}(p^{\prime0}M_{\sigma})^{2}\frac{1}{\vec{k}_{m}^{2}}\label{k2}\\
 & \sum_{r=1}^{n-1}\vec{k}_{r}^{2}\frac{\partial}{\partial(2i\vec{p^{\prime}}\cdot\vec{k}_{r})}\left[\frac{1}{\vec{p^{\prime}}\cdot\vec{k}_{1}}\frac{1}{\vec{p^{\prime}}\cdot\vec{k}_{2}}\dots\frac{1}{\vec{p^{\prime}}\cdot\vec{k}_{n-1}}\right](2\pi)^{4}\delta^{3-2\epsilon}(\vec{\Delta}+\vec{k}_{1}+\dots\vec{k}_{n}).\nonumber 
\end{align}
Due to (\ref{impact}) the contribution to the amplitude due to terms proportional to $k^2$ is concentrated around $\vec{b}=0$ and are therefore disregarded for our purposes due to (\ref{bigimpact}). In (\ref{NE}) start by considering the sums with different indices
\begin{eqnarray}
&&\sum\limits_{\substack{m, m_1=1\\ m\neq m_1}}^n k_m\cdot k_{m_1}\left(\theta(\xi_m-\xi_{m_1})(\nu-\xi_m)+ \theta(\xi_{m_1}-\xi_{m})(\nu-\xi_{m_1}) \right)+\nonumber\\&&\sum\limits_{\substack{m_2, m_3=1\\m_2\neq m_3}}^n k_{m_2}\cdot k_{m_3}(\theta(\tilde{\xi}_{m_2}-\tilde{\xi}_{m_3})(\nu_1-\tilde{\xi}_{m_2})+ \theta(\tilde{\xi}_{m_3}-\tilde{\xi}_{m_2})(\nu_1-\tilde{\xi}_{m_3})).\nonumber\\
\end{eqnarray}
We are going to prove that since $M_\sigma>> E_\phi$ the only needed sum is
\begin{eqnarray}
\sum_{m, m_1=1, m\neq m_1}^n k_m\cdot k_{m_1}\left(\theta(\xi_m-\xi_{m_1})(\nu-\xi_m)+ \theta(\xi_{m_1}-\xi_{m})(\nu-\xi_{m_1}) \right).
\end{eqnarray}
Let $m,m_1\in \{1,\dots n\}$, $m\neq m_1$ and consider the integral
\begin{eqnarray}
&&\int_0^{\nu} d\xi_m d\xi_{m_1}\exp[-2ip’\cdot k_m(\nu-\xi_m)-2ip’\cdot k_{m_1}(\nu-\xi_{m_1})]\nonumber\\&&\quad \times[\theta(\xi_m-\xi_{m_1})(\nu-\xi_m)+\theta(\xi_{m_1}-\xi_m)(\nu-\xi_{m_1}) ]=\nonumber\\&&  -\frac{1}{(2ip’\cdot k_m)(2ip’\cdot k_{m_1})(2ip’\cdot(k_m+k_{m_1}))}\left[1-e^{(-2ip’\cdot(k_m+k_{m_1})\nu)}\right] \label{1} \\
&& + \quad \frac{\nu}{(2ip’\cdot k_m)(2ip’\cdot k_{m_1})}e^{(-2ip’\cdot(k_m+k_{m_1})\nu)} \quad + \label{2}\\
&&\frac{2ip’\cdot(k_m+k_{m_1})}{(2ip’\cdot k_m)^2 (2ip’\cdot k_{m_1})^2}\left(1-e^{(-2ip’\cdot k_m \nu)}\right)\left(1-e^{(-2ip’\cdot k_{m_1} \nu)}\right)\label{3}
\end{eqnarray} 

The contribution of the term (\ref{1}) to the $i\mathcal{T}$ matrix amounts to 
\begin{eqnarray}
&&i\mathcal{T}_{NE}^{2a}\equiv\lim\limits_{\substack {p^2\rightarrow 0\\ p’^2\rightarrow 0}} p^2p^{\prime 2}(2\pi)^4\sum_{n=2}^{\infty}\frac{i^n\kappa^{2n}}{n!}\int_0^{+\infty}d\nu 
e^{i\nu(p’^2+i\epsilon)}\frac{(E_\phi M_\sigma)^{2n}}{(2\pi)^{4n}}\sum\limits_{\substack{\tilde{m}, m_1=1\\ \tilde{m}\neq m_1}}^n\nonumber\\&&
\int\frac{d^{3-2\epsilon} k_1}{\vec{k}_1^2}\dots \frac{d^{3-2\epsilon} k_n}{\vec{k}_n^2}\delta^{3-2\epsilon} (\vec{k}_1+\dots+ \vec{k}_n +\vec{\Delta}) \vec{k}_{\tilde{m}}\cdot \vec{k}_{m_1}\frac{i^n(-2\pi i)^{n-1}}{(2M_\sigma)^{n-1}}\nonumber\\&&
\prod\limits_{\substack{m=1\\m\neq \tilde{m}, m_1}}^n \frac{1-\exp(2i\nu\vec{p}’\cdot \vec{k}_m)}{-2i\vec{p}’\cdot\vec{k}_m} 
\frac{[1-\exp(2i\vec{p}’\cdot(\vec{k}_{\tilde{m}}+\vec{k}_{m_1})\nu)]}{(-2i\vec{p}’\cdot \vec{k}_{\tilde{m}})(-2i\vec{p}’\cdot \vec{k}_{m_1})(-2i\vec{p}’\cdot(\vec{k}_{\tilde{m}}+\vec{k}_{m_1}))}\nonumber\\
\label{DEF}
\end{eqnarray}

Once again by the use of (\ref{eik1}), (\ref{eik2}), (\ref{eik22}) we obtain
\begin{eqnarray}
&&\lim\limits_{\substack {p^2\rightarrow 0\\ p’^2\rightarrow 0}} p^2p^{\prime 2}\delta(k^z_1+\dots + k^z_n)
\prod\limits_{\substack{m=1\\m\neq \tilde{m}, m_1}}^n \frac{1-\exp(2i\nu\vec{p}’\cdot \vec{k}_m)}{-2i\vec{p}’\cdot\vec{k}_m}\times\nonumber\\&& 
\frac{[1-\exp(2i\vec{p}’\cdot(\vec{k}_{\tilde{m}}+\vec{k}_{m_1})\nu)]}{(-2i\vec{p}’\cdot \vec{k}_{\tilde{m}})(-2i\vec{p}’\cdot \vec{k}_{m_1})(-2i\vec{p}’\cdot(k_{\tilde{m}}+k_{m_1}))}=\nonumber\\&&
\frac{(-2\pi i)^{n-2}}{(2E_\phi)^{n-2}}\delta(k_1^z)\dots \delta(k_{n-2}^z)\delta(k^z_{\tilde{m}}+ k^z_{m_1})
\end{eqnarray}
and replacing into (\ref{DEF}) we obtain 
\begin{eqnarray}
&&i\mathcal{T}_{NE}^{2a}=i\sum_{n=2}^\infty\left(\frac{i\kappa^2 E_\phi M_\sigma}{16\pi^2}\right)^n\frac{1}{(n-2)!}\frac{M_\sigma}{2\pi}\left(\int\frac{d^{2-\epsilon} k_m}{\vec{k}_m^{\perp 2}}\right)^{n-2}
\nonumber\\ && \int\frac{d^{3-\epsilon} k_{\tilde{m}}}{\vec{k}_{\tilde{m}}^2}\frac{d^{3-2\epsilon} k_{m_1}}{\vec{k}_{m_1}^2}
 \frac{\vec{k}_{\tilde{m}}\cdot \vec{k}_{m_1}}{k_{\tilde{m}}^z k_{m_1}^z}\delta(k_{\tilde{m}}^z+k_{m_1}^z)(2\pi)^2\delta^2(\vec {k}^\perp_1+ \dots+\vec {k}^\perp_n +\vec{\Delta})\nonumber\\
\label{NED}
\end{eqnarray}
where $\frac{1}{2(n-2)!}=\frac{1}{n!}\times {{n}\choose{2}}$ comes from the sum over the equivalent combinations of two elements among a set of $n$ elements.
The contribution from (\ref{2}) to the next-to-eikonal approximation of the scattering amplitude (\ref{NE}) is the same as (\ref{NED}) as it can be understood by the replacement into (\ref{2})
\begin{equation}
-\nu\delta(k^z_{m} +k^z_{m_1})=\delta(k^z_{m} +k^z_{m_1})\frac{1-e^{2i\nu E_\phi (k^z_{m} +k^z_{m_1})}}{2i\nu E_\phi (k^z_{m} +k^z_{m_1})}.
\end{equation} 
as consequence of the eikonal identities. Due to the $\delta(k^z_{m} +k^z_{m_1})$ the exponential factor in (\ref{2}) amounts to one.
Concerning the contribution of  (\ref{3}) to the Fourier transform of the amplitude (\ref{NE}) it vanishes because each ladder diagram is proportional to
\begin{equation}
\int_{-\infty}^{+\infty}\frac{d k_1^z}{k_1^{z 2} +k_1^{\perp 2}}\frac{\delta(k_1^z)}{k_1^z}\int_{-\infty}^{+\infty}\frac{d k_2^z}{k_2^{z 2} +k_2^{\perp 2}}{\delta(k_2^z)}=0.
\end{equation}
By taking the Fourier transform to the impact parameter space we obtain 
\begin{eqnarray}
&&i\widetilde{\mathcal{T}}(\vec{b}^\perp)_{NE}=\int \frac{d^{2-\epsilon}\vec{\Delta}}{(2\pi)^2} e^{i\vec{\Delta}\cdot\vec{b}^\perp} i\mathcal{T}_{NE}^{2a}=\nonumber\\&&i\int \frac{d^{2-\epsilon}\vec{\Delta}}{(2\pi)^2} e^{i\vec{\Delta}\cdot\vec{b}^\perp} \sum_{n=2}^\infty\left(\frac{i\kappa^2 E_\phi M_\sigma}{16\pi^2}\right)^n\frac{2}{(n-2)!}\frac{M_\sigma}{2\pi}\left(\int\frac{d^{2-2\epsilon} k_m}{\vec{k}_m^{\perp 2}}\right)^{n-2}
\nonumber\\ && \int\frac{d^{3-2\epsilon} k_{\tilde{m}}}{\vec{k}_{\tilde{m}}^2}\frac{d^{3-2\epsilon} k_{m_1}}{\vec{k}_{m_1}^2}
 \frac{\vec{k}_{\tilde{m}}\cdot \vec{k}_{m_1}}{k_{\tilde{m}}^z k_{m_1}^z}\delta(k_{\tilde{m}}^z+k_{m_1}^z)(2\pi)^2\delta^{2-2\epsilon}(\vec {k}^\perp_1+ \dots+\vec {k}^\perp_n +\vec{\Delta} )\nonumber\\&&
= \quad 2i (s-M_\sigma^2)\sum_{n=2}^\infty \frac{(i\chi_0(\vec{b}^\perp))^{n-2}}{(n-2)!}\tilde{\chi}(\vec{b}).
\label{TNE}
\end{eqnarray}
Here $(s-M_\sigma^2)=2E_\phi M_\sigma$, the exponentiated eikonal phase $\chi_0(\vec{b}^\perp)$ with its definition in (\ref{fase}) is recovered. Moreover
\begin{equation}
\tilde{\chi}(\vec{b})=\frac{ \kappa^4 E_\phi M_\sigma^2}{1024\pi^5}\int dk_1^z d^{2-2\epsilon}k_1^\perp d^{2-2\epsilon} k_2^\perp \frac{e^{-i\vec{b}^\perp\cdot \vec{k}_1^\perp}}{k_1^{\perp 2}+k_1^{z 2}}\frac{e^{-i\vec{b}^\perp\cdot \vec{k}_2^\perp}}{k_2^{\perp 2}+k_1^{z 2}}\frac{\vec{k}_1^\perp\cdot \vec{k}_2^\perp -k_1^{z 2}}{k_1^{z 2}}.
\label{chiNE}
\end{equation}
By following the same derivation of \cite{Sterman} we obtain for (\ref{chiNE}) 
\begin{equation}
\tilde{\chi}(\vec{b}^{\perp})=\frac{\kappa^{4}E_{\phi}M_{\sigma}^{2}}{1024\pi^{5}}\left[2\pi b^{6\epsilon-1}\frac{\Gamma\left(\frac{1}{2}-2\epsilon\right)\Gamma\left(\frac{1}{2}-\epsilon\right)}{\Gamma\left(-\epsilon\right)}\right],\label{eq:Chibperp}
\end{equation}
with $b=|\vec{b}^\perp|$ and for $\epsilon\rightarrow 0$ meaning in four spacetime dimensions we get
\begin{equation}
\tilde{\chi}(\vec{b}^\perp)=0,
\end{equation}
in agreement with \cite{DIV}, \cite{DIV1} as well as \cite{Sterman}, \cite{BB}. It means that in four dimensions the next-to-eikonal correction of (\ref{TNE}) starts at higher orders. However if in this approximation $e^{i\chi_0}$ is modified by a phase also in this functional context is not clear to us.

\section{Multi-gravitons exchange beyond ladder\label{SEI}}
We consider the next to eikonal correction to the amplitude coming from the the insertion of a seagull interaction and a trilinear graviton vertex. Consider first of all the insertion of one seagull interaction vertex inside the sum of the ladder diagrams. This is achieved by expanding the spin zero matter action (\ref{matter}) 
untill the quartic interaction, the so called seagull terms \cite{donnoghue}, 
\begin{align}
S & =-\frac{1}{2}\int d^{4}x\phi(x)\left[\partial^{2}+m^{2}-\kappa h^{\mu\nu}\partial_{\mu}\partial_{\nu}+\frac{\kappa}{2}\eta_{\mu\nu}h^{\mu\nu}\partial^{2}+\frac{\kappa}{2}\eta^{\mu\nu}h_{\mu\nu}m^{2}\right.\nonumber \\
 & \quad-\frac{1}{4}\kappa^{2}\partial_{\alpha}(h^{\mu\nu}h_{\mu\nu})\partial^{\alpha}+\frac{\kappa^{2}}{8}\partial_{\alpha}(h_{\mu}^{\mu})^{2}\partial^{\alpha}+\frac{\kappa^{2}}{8}(h_{\mu}^{\mu})^{2}\partial^{2}\nonumber \\
 & \quad+\frac{k^{2}}{8}m^{2}(h_{\mu}^{\mu})^{2}-\frac{k^{2}}{2}\partial_{\alpha}(h_{\mu}^{\mu}h^{\alpha\beta})\partial_{\beta}-\frac{\kappa^{2}}{2}\partial_{\alpha}(h_{\mu}^{\mu}h^{\alpha\beta})\partial_{\beta}\nonumber \\
 & \quad\left.-\frac{\kappa^{2}}{2}h_{\mu}^{\mu}h^{\alpha\beta}\partial_{\alpha}\partial_{\beta}+\:\kappa^{2}h^{\mu\lambda}h_{\lambda}^{\nu}\partial_{\mu}\partial_{\nu}+\kappa^{2}\partial_{\mu}(h^{\mu\lambda}h_{\lambda}^{\nu})\partial_{\nu}\right]\phi(x).\label{eq:seagul}
\end{align}
By following the same steps as in the section (\ref{DUE}) we arrive to the following Cauchy problem
\begin{equation}
\begin{cases} 
i\frac{\partial Y}{\partial\nu}(x,\nu)=\\\left[\partial^2-2i \ell_\alpha\partial^\alpha +\kappa \ell_\mu \ell_\beta h^{\mu\beta}(x)+\frac{\kappa^2}{2}h^\mu\,_\mu h^{\alpha\beta}\ell_\alpha\ell_\beta- \kappa^2 h^{\alpha\lambda}h_{\lambda}\,^\beta\ell_\alpha\ell_\beta  \right]Y(x,\nu) \\ 
Y(x,0)=1.
\end{cases}
\label{Cauchy2}
\end{equation}
By following the same steps as in (\ref{Cauchy1}) the corresponding solution is
\begin{eqnarray}
&&Y(x,\nu)=\exp\left[-i\kappa \ell_\mu \ell_\beta \int_0^\nu h^{\mu\beta}\left(x+2\ell(\nu’-\nu)+\int_{\nu’}^{\nu} \frac{\delta}{\delta t(\xi)}d\xi\right)d\nu’ \right.\nonumber\\&&\left. -i\frac{\kappa^2}{2}\ell_\alpha\ell_\beta\int_0^\nu h^{\mu}\,_\mu h^{\alpha\beta}\left(x+2\ell(\nu’-\nu)+\int_{\nu’}^\nu\frac{\delta}{\delta t(\xi)}d\xi\right)d\nu’\right.\nonumber\\&&\left. 
+i\kappa^2 \ell_\alpha\ell_\beta\int_0^\nu h^{\alpha\lambda} h_{\lambda}\,^\beta\left(x+2\ell(\nu’-\nu)+\int_{\nu’}^\nu\frac{\delta}{\delta t(\xi)}d\xi\right)d\nu’\right]
\nonumber\\ &&\left.\exp\left(i\int_0^\nu t^2(\xi)d\xi \right) \right |_{t=0}
\label{sol2}
\end{eqnarray} 
The contribution to $Y(x,\nu)$ corresponding to the sum of $n$ gravitons exchange diagrams with a single seagull in the amplitude (\ref{Gc}) for the propagation of the light scalar by emitting gravitons in the straight line approximation is
\begin{eqnarray}
&&<p’|G^c(x,y|h)|p>=\left(-i\frac{\kappa^2}{2}\right) \lim\limits_{\substack{p^2\rightarrow 0 \\ p^{\prime 2}\rightarrow 0}}p^2 p^{‘2}(2\pi)^4 \left(p^{\prime\alpha}p^{\prime\beta}\eta^{\gamma\delta}-2p^{\prime\gamma}p^{\prime\beta}\eta^{\delta\alpha}\right)\nonumber\\&&\sum_{n=0}^{+\infty}\frac{(-i)^n\kappa^n}{n!}\int_0^{+\infty}d\nu\, e^{i\nu(p^{’2}+i\epsilon)}\int\frac{d^4 \bar{k} d^4\underline{k}}{(2\pi)^8}\int\frac{d^4 k_1\dots d^4k_n}{(2\pi)^{4n}}\int_0^\nu d\nu’ \int_0^\nu d\xi_1\nonumber\\&&\dots d\xi_n \delta^4(p’-p-k_1-\dots -k_n-\bar{k}-\underline{k}) e^{-\sum\limits_{m=1}^n 2ip’\cdot k_m(\nu-\xi_m)}
\nonumber\\&& e^{-2ip’\cdot (\bar{k}+\underline{k})(\nu-\nu’)} p^{\prime\mu_1}p^{\prime\beta_1}\hat{h}_{\mu_1\beta_1}(k_1)\dots p^{\prime\mu_n}p^{\prime\beta_n}\hat{h}_{\mu_n\beta_n}(k_n)\hat{h}_{\gamma\delta}(\bar{k})\hat{h}_{\alpha\beta}(\underline{k}).\nonumber\\
\end{eqnarray}
For the scattering amplitude of our high-energy process with the insertion of one seagull interaction we need for the expansion of the heavy line propagator at least two emitted gravitons amounting
\begin{eqnarray}
&&  \lim\limits_{\substack {q^2\rightarrow M_\sigma^2\\ q’^2\rightarrow M_\sigma^2}}(q^2-M_\sigma^2)(q^2-M_\sigma^2)(2\pi)^4\sum_{r=2}^{+\infty}\frac{(-i)^r\kappa^r}{r!}\int_0^{+\infty}d\nu_1 e^{i\nu_1(q’^2-M_\sigma^2+i\epsilon)}\nonumber\\&&\int\frac{d^4 \tilde{k}_1\dots d^4\tilde{k}_r}{(2\pi)^{4r}}\int_0^{\nu_1} d\xi_1\dots\nonumber\int_0^{\nu_1} d\xi_n \delta^4(q’-q+\tilde{k}_1+\dots \tilde{k}_r)\nonumber\\&& e^{-\sum\limits_{m=1}^r 2i q’\cdot \tilde{k}_m(\nu_1-\xi_m)} q’^{\mu_1}q’^{\beta_1}\hat{h}_{\mu_1\beta_1}(\tilde{k}_1)\dots q’^{\mu_r}q’^{\beta_r}\hat{h}_{\mu_r\beta_r}(k_r).
\label{HL}
\end{eqnarray}
The Wick’s contractions performed by following the formula (\ref{T-matrix}) implies the need of the following matrix element 
\begin{eqnarray}
&&\sum\limits_{n=0}^{+\infty}\sum\limits_{r=2}^{+\infty}(-i)^{n+2}(-i)^r\frac{\kappa^{n+2}\kappa^r}{2\,n! r!}\int\frac{d^4 k_1\dots d^4k_n d^4 \bar{k} d^4 \underline{k}}{(2\pi)^{4(n+2)}}\int \frac{d^4 \tilde{k}_1\dots d^4 \tilde{k}_r}{(2\pi)^{4r}}
\nonumber\\&&
\delta^4(p’-p-k_1-\dots-k_n-\underline{k}-\bar{k})\delta^4(q’-q-\tilde{k}_1-\dots-\tilde{k}_r)\times\nonumber\\&&<p^{\prime\mu_1}p^{\prime\beta_1}\hat{h}_{\mu_1\beta_1}(k_1)\dots p^{\prime\mu_n}p^{\prime\beta_n}\hat{h}_{\mu_n\beta_n}(k_n)(p^{\prime\alpha}p^{\prime\beta}\eta^{\rho\sigma}-2p^{\prime\rho}p^{\prime\beta}\eta^{\sigma\alpha})\nonumber\\&&\hat{h}_{\rho\sigma}(\bar{k})\hat{h}^{\alpha\beta}(\underline{k})q^{\prime\tilde{\mu}_1}q^{\prime\tilde{\beta}_1}\hat{h}_{\tilde{\mu}_1\tilde{\beta}_1}(\tilde{k}_1)\dots q^{\prime\tilde{\mu}_r}q^{\prime\tilde{\beta}_r}\hat{h}_{\tilde{\mu}_r\tilde{\beta}_r}(\tilde{k}_r)>,
\end{eqnarray}
where by the prescription (\ref{T-matrix}) the self-energy contributions are not included. We get for such matrix element
\begin{eqnarray}
&&\sum\limits_{r=2}^{+\infty}\frac{(-i)^{2r}\kappa^{2r}}{2(r-2)!}\int \frac{d\tilde{k}_1\dots d^4\tilde{k}_r}{(2\pi)^{4r}}(2\pi)^4\delta^4(q’-q-\tilde{k}_1-\dots-\tilde{k}_r)\frac{M_\sigma^4 E_\phi^2}{\tilde{k}_1^2\tilde{k}_2^2}\nonumber\\&& 
\prod\limits_{m=3}^r p^{\prime\mu_m}p^{\prime\beta_m}q^{\prime\tilde{\mu}_m}q^{\prime\tilde{\beta}_m}\frac{i}{2\tilde{k}_m^2} (\eta_{\mu_m\tilde{\mu}_m}\eta_{\beta_m\tilde{\beta}_m}+ \eta_{\mu_m\tilde{\beta}_m} \eta_{\beta_m\tilde{\mu}_m}-\eta_{\mu_m\beta_m} \eta_{\tilde{\beta}_m\tilde{\mu}_m} ).\nonumber\\
\end{eqnarray}
The leading contribution to the amplitude including the non-linear seagull interaction amounts to
\begin{eqnarray}
&&i\mathcal{T}^{NL}(p,p’,q,q’)_{SG}=\nonumber\\&&\frac{i}{2} \lim_{p_i^2\rightarrow m_i^2} p^2 p’^2 (q^2-M_\sigma^2)(q’^2-M_\sigma^2)\sum_{r=2}^{+\infty}\frac{(-i)^{2r}\kappa^{2r}}{(r-2)!}\int_0^{+\infty}d\nu e^{i\nu(p’^2+i\epsilon)}\nonumber\\&&
\int_0^{+\infty}d\nu_1 e^{i\nu_1(q’^2-M_\sigma^2 +i\epsilon)}\frac{1}{(2\pi)^{4r}}\int d^4 \bar{k} d^4 \underline{k}\frac{M_\sigma^4 E_\phi^2}{\bar{k}^2\underline{k}^2}\left(\prod_{m=3}^{r}\int d^4 k_m\frac{i}{k_m^2} E_\phi^2 
M_\sigma^2\right)\nonumber\\&& 
(2\pi)^4
\delta^4(p’-p-k_1-\dots-k_n-\bar{k}-\underline{k})\int_0^\nu\exp\left[-\sum_{m=1}^{r-2}2ip’\cdot k_m(\nu-\xi_m)\right]\nonumber\\&&d\xi_1\dots d\xi_{r-2}\int_0^{\nu_1}\exp\left[\sum_{\tilde{m}=1}^{r}2iq’\cdot k_{\tilde{m}}(\nu_1-\tilde{\xi}_{\tilde{m}})\right]d\tilde{\xi}_1\dots d\tilde{\xi}_{r}\times\nonumber\\&&\int_0^\nu d\nu’ e^{-2ip’\cdot (\bar{k}+\underline{k})(\nu-\nu’)}
\end{eqnarray}
By applying the eikonal identities of (\ref{eik1}), (\ref{eik2}), (\ref{eik3}) we obtain
\begin{eqnarray}
&&i\mathcal{T}^{NL}(p,p’,q,q’)_{SG}=\nonumber\\&&-\frac{1}{2} \lim_{p_i^2\rightarrow m_i^2} p^2 p’^2\sum_{r=2}^{+\infty}\frac{(-i)^{2r}\kappa^{2r}}{(r-2)!}\int_0^{+\infty}d\nu\,\frac{e^{i\nu(p’^2+i\epsilon)}}{(2\pi)^{4r}}
\int d^4 \bar{k} d^4 \underline{k}\frac{M_\sigma^4 E_\phi^2}{\bar{k}^2\underline{k}^2}\nonumber\\&&\left(\prod_{m=3}^{r}\int d^4 k_m\frac{i}{k_m^2} E_\phi^2 M_\sigma^2\right)\int_0^\nu\exp\left[-\sum_{m=1}^{r-2}2ip’\cdot k_m(\nu-\xi_m)\right]d\xi_1\dots d\xi_{r-2}\nonumber\\&&\int_0^\nu d\nu’ e^{-2ip’\cdot (\bar{k}+\underline{k})(\nu-\nu’)}
i^r\frac{(-2\pi i)^{r-1}}{(2M_\sigma)^{r-1}}
\delta(k^0_1)\dots \delta(k^0_r)\times\nonumber\\&&(2\pi)^4\delta^3(\vec{k}_1+\dots+\vec{k}_n+\vec{\bar{k}}+\vec{\underline{k}}+\vec{\Delta})
\end{eqnarray}
To apply the eikonal identities along the $z$-axis consider 
\begin{align}
 & \lim_{p_{i}^{2}\rightarrow m_{i}^{2}}p^{2}p\text{\textquoteright}^{2}\delta(k_{1}^{z}+\dots k_{r-2}^{z}+\bar{k}^{z}+\underline{k}^{z})\int_{0}^{+\infty}d\nu e^{i\nu(p\text{\textquoteright}^{2}+i\epsilon)}\times\nonumber \\
 & \frac{1-e^{2i\nu E_{\phi}(\bar{k}^{z}+\underline{k}_{z})}}{-2iE_{\phi}(\bar{k}^{z}+\underline{k}_{z})}\prod_{m=3}^{r}\frac{1-e^{2i\nu E_{\phi}k_{m}^{z}}}{-2iE_{\phi}k_{m}^{z}}\:=\nonumber \\
 & i^{r}\frac{(-2\pi i)^{r-2}}{(2E_{\phi})^{r-2}}\delta(k_{1}^{z})\dots\delta(k_{r-2}^{z})\delta(\bar{k}^{z}+\underline{k}^{z}).\label{eq:eikide}
\end{align}
The contribution to the amplitude becomes 
\begin{align}
 & i\mathcal{T}^{NL}(p,p\text{\textquoteright},q,q\text{\textquoteright})_{SG}=\nonumber \\
 & \sum\limits _{r=2}^{+\infty}\frac{i}{(r-2)!}\left(\frac{i\kappa^{2}M_{\sigma}E_{\phi}}{4(2\pi)^{2}}\right)^{r-2}\left[\frac{\kappa^{4}M_{\sigma}^{3}E_{\phi}^{2}}{32\pi^{3}}\int\frac{d^{3}\bar{k}d^{3}\underline{k}}{\vec{\bar{k}}^{2}\vec{\underline{k}}^{2}}\delta(\bar{k}^{z}+\underline{k}^{z})\right]\nonumber \\
 & \int\prod\limits _{m=3}^{r}d^{2}\tilde{k}_{m}^{\perp}\frac{1}{(\tilde{k}_{m}^{\perp})^{2}}\delta^{2}(\vec{\Delta}+\vec{k}_{1}+\dots+\vec{k}_{n}+\vec{\bar{k}}+\vec{\underline{k}}).\label{seag}
\end{align}
The contribution beyond the ladders coming from the three-linear vertex of gravitons is coming from (\ref{amplitude}) implying (\ref{T-matrix}) with the insertion of the three linear gravitons vertex. For that calculation the functional methods are totally equivalent to diagrammatic ones already described \cite{Sterman} for which we just quote the result for the contribution to the scattering amplitude
\begin{eqnarray}
&&i\mathcal{T}^{NL}(p,p’,q,q’)_{3g}\equiv -\sum\limits_{r=2}^{+\infty}\frac{i}{(r-2)!}\left(\frac{i\kappa^2 M_\sigma E_\phi}{4(2\pi)^2}\right)^{r-2}\left(\frac{\kappa^4 M_\sigma^3 E_\phi^2}{64\pi^3}\right)\times\nonumber\\&&\int\frac{d^3 \bar{k} d^3\underline{k}}{\vec{\bar{k}}^2\vec{\underline{k}}^2}\frac{\bar{k}_z^2}{\vec{\Delta}^2}\delta(\bar{k}^z+\underline{k}^z)
\int\prod\limits_{m=3}^r d^2\tilde{k}^\perp_m\frac{1}{(\tilde{k}^\perp_m)^2}\delta^2(\vec{\Delta}+\vec{k}_1+\dots+\vec{k}_n+\vec{\bar{k}}+\vec{\underline{k}}).\nonumber\\
\label{3G}
\end{eqnarray}
The total next-to-eikonal correction for the inclusion of non linear interactions is obtained by summing (\ref{seag}) plus (\ref{3G}) amounting to 
\begin{align}
i\widetilde{\mathcal{T}}^{NL}(\vec{b}^{\perp}) & \equiv i\widetilde{\mathcal{T}}^{NL}(\vec{b}^{\perp})_{SG}+i\widetilde{\mathcal{T}}^{NL}(\vec{b}^{\perp})_{3g}\nonumber \\
 & =2i(s-M_{\sigma}^{2}))\sum_{n=2}^{\infty}\frac{(i\chi_{0}(\vec{b}^{\perp}))^{n-2}}{(n-2)!}\tilde{\chi}_{1}(\vec{b}).\label{totalnte}
\end{align}
with 
\begin{align}
\tilde{\chi}_{1}(\vec{b}) & =\frac{\kappa^{4}M_{\sigma}^{2}E_{\phi}}{128\pi^{3}}\int\frac{d^{2}\Delta}{(2\pi)^{2}}e^{i\vec{b}^{\perp}\cdot\vec{\Delta}}\int\frac{d^{3}\bar{k}d^{3}\underline{k}}{\vec{\bar{k}}^{2}\vec{\underline{k}}^{2}}\left(1+\frac{\bar{k}_{z}^{2}}{2\vec{\Delta}^{2}}\right)\delta(\bar{k}^{z}+\underline{k}^{z})\nonumber \\
 & =\frac{15\kappa^{4}M_{\sigma}^{2}E_{\phi}}{4096\pi\left|\vec{b}^{\perp}\right|}\label{Chi1b}
\end{align}
where standard Feynman parametrization technique has been used to find the above result which agrees with \cite{DIV}, \cite{DIV1} as well as \cite{Sterman}, \cite{BB}.
  
\section{Conclusions and research perspectives\label{SETTE}}
In this paper we have analyzed the eikonal limit for the gravitational scattering of a high-energy or massless scalar particle by a very heavy scalar. The eikonal expansion has been performed by using the Fradkin’s representation of the scattering amplitudes and we have shown that by those functional methods we get a totally consistent eikonal expansion with a resummation of diagrams equivalent to the one performed in \cite{Sterman}. Moreover, we have shown that the exponentiation of the leading eikonal applies as well for the first non-leading power in the energy of the light particle. Working at leading power in the heavy particle mass, we expanded the light particle propagator to next-to-eikonal power and included gravitational interactions of comparable size, finding power corrections suppressed by a single power of $\frac{\Delta}{E_\phi}$, or $\frac{E_\phi}{b}$ in the impact parameter space, with respect to the leading eikonal term. The comparable terms are based on one-loop diagrams of order $\kappa^4$. In our analysis corrections are factors leaving leading-power exponentiation unaffected. The consistency with the exponentiation of the power corrections themselves in our functional formalism has not been yet explored. Next-to-eikonal corrections vanish in four dimensions, the way of possible exponentiation in our representation is a topic for future researches. Our calculations methods could be also useful for the forward scattering of a gravitational wave from a black hole. For further research proposal on the line of our study we remark that in a recent article \cite{bernlast} the conservative two bodies dynamics for spinless compact objects described by a four-point amplitude truncated to classical order with the matter poles expanded about the momentum components along $z$ was considered. This kind of expansion is automatically done in the Fradkin representation of the scalar propagator meaning in the formula like (\ref{T1}) after the applications of the eikonal identities  (\ref{eik}), (\ref{eik1}), (\ref{eik2}), (\ref{eik22}) and (\ref{eik3}). From the eikonalization of the scattering amplitude for the small momentum transfer an amplitude-action relation arises, the determination of the radial action by using the Fradkin representation is something that could be efficiently achieved, even if a treatment Lagrangian independent should be introduced and this will be left for future studies. The methods adopted in our paper could also illuminate the contributions to two body Hamiltonian from an infinite family of tidal operators, that was started in the paper \cite{bernbl}. The functional approach to scattering amplitudes in the functional representation could indeed be optimized to include insertions of powers of the position-space magnetic and electric components of the linearized Weyl tensor contracted with a point particle stress tensor. Another direction could be to consider, like in the reference \cite{trava}, effective theories of gravity where in addition to the Einstein-Hilbert term non-minimal couplings of the type $R^3$, $R^4$ and $FFR$ could be included, being $F$ the photon field strenght. By using the Fradkin representation for the scattering of gravitons and photons of frequency $\omega$ off heavy scalars of mass $m$ in the limit $m>>\omega>>|\vec{q}|$, where $\vec{q}$ is the momentum transfer, it would be interesting to see in this context the eikonal and next-to-eikonal limit of this amplitude by functional techniques, and compare with the analogous exponentiation in \cite{ACV1}, \cite{ACV2}. In the paper of \cite{monteiro} it is proved that the quantum state setup by a particle is a coherent state fully determined due to an eikonal type exponentiation, it would also be interesting to discuss, by using the tools introduced in this paper, how the coherent state changes by including non-linear interactions and also next-to-eikonal approximations.

\end{document}